\documentstyle[pra,aps,amssymb,psfig,twocolumn]{revtex}

\newcommand{\mbb}[1]{\mbox{\boldmath $#1$}}

\begin{document}

\draft

\title{Entanglement transformation at absorbing and amplifying
four-port devices}
\author{S. Scheel$^1$, L. Kn\"{o}ll$^1$, T. Opatrn\'{y}$^{1,2}$, 
and D.-G. Welsch$^1$}
\address{$^1$Theoretisch-Physikalisches Institut,
Friedrich-Schiller-Universit\"at Jena, Max-Wien-Platz 1, D-07743 Jena,
Germany\\
$^2$Department of Theoretical
Physics, Palack\'{y} University, Svobody 26, 771~46 Olomouc, Czech Republic} 

\date{May 15, 2000}
\maketitle

\begin{abstract}

Dielectric four-port devices play an important role in
optical quantum information processing. Since for
causality reasons the permittivity is a complex
function of frequency, dielectrics are typical
examples of noisy quantum channels, which 
cannot preserve quantum coherence. To study
the effects of quantum decoherence, we start from
the quantized electromagnetic field in an arbitrary
Kramers--Kronig dielectric of given complex
permittivity and construct the transformation
relating the output quantum state to the
input quantum state, without placing restrictions
on the frequency. We apply the formalism
to some typical examples in quantum communication.
In particular we show that for Fock-entangled 
qubits the Bell-basis states $|\Psi^\pm\rangle$
are more robust against decoherence than the states
$|\Phi^\pm\rangle$.
\end{abstract}



\section{Introduction}
\label{intro}

Quantum communication schemes widely use
dielectric four-port devices as basic elements
for constructing optical quantum channels. A typical 
example of such a device is a beam splitter as
a basic element not only for classical interference
experiments but also for implementing quantum 
interferences. Another example is an optical fiber,
which can be regarded as a dielectric four-port device 
that essentially realizes transmission of light over 
longer distances. 

Dielectric matter is commonly described
in terms of the (spatially varying) permittivity
as a complex function of frequency, whose real and
imaginary parts are related to each other by the
Kramers--Kronig relations. Since the appearance of
the imaginary part (responsible for absorption and/or
amplification) is unavoidably associated with
additional noise, dielectric devices are typical
examples of noisy quantum channels. Using
them for generating or processing entangled
quantum states of light, e.g., in quantum
teleportation or quantum cryptography, 
the question of quantum decoherence arises.  

In order to study the problem, quantization of the 
electromagnetic field in the presence of dielectric media
is needed. For absorbing bulk material, a consistent
formalism is given in \cite{Huttner92}, using 
the Hopfield model of a dielectric \cite{Hopfield58}.
A method of direct quantization of Maxwell's equations 
with a phenomenologically introduced permittivity is given
in \cite{Gruner96a}. It replaces the familiar mode
decomposition of the electromagnetic field with 
a source-quantity representation, expressing the field in terms 
of the classical Green function and the fundamental variables
of the composed system. The method has the benefit 
of being independent of microscopic models of the medium 
and can be extended to arbitrary inhomogeneous dielectrics 
\cite{Ho98,Scheel98}. All relevant information about the
medium are contained in the permittivity (and the resulting
Green function), and quantization is performed by the
association of bosonic quantum excitations with the
fundamental variables. 

Quantization of the phenomenological
Maxwell field is especially well suited for deriving the
input-output relations of the field 
\cite{Matloob95,Matloob96,Gruner96b,Barnett96} on the 
basis of the really observed transmission and absorption coefficients.
In particular, there is no need to introduce 
artificial replacement schemes. 
Applications to low-order correlations in two-photon
interference effects have been given \cite{Gruner96b,Gruner97,Barnett98}.

The formalism has also been extended to amplifying
media \cite{Scheel98,Jeffers96OC}. The resulting 
input--output relations for amplifying beam splitters have been 
used to compute first- and second-order moments of photo counts 
\cite{Jeffers93} and normally ordered Poynting vectors 
\cite{Artoni98}. Further, propagation of squeezed radiation 
through amplifying or absorbing multiport devices
has been considered \cite{Beenakker99}.

For the study of entanglement, however, knowledge of
some moments and correlations is not enough. 
In particular, to answer the question as
to whether or not a bipartite quantum state is
separable and to calculate the degree of entanglement
of a nonseparable state, the complete information on the  
state is required in general.

Recently we have presented closed formulas for calculating
the output quantum state from the input quantum state \cite{Knoll99}, 
using the input--output relations for the field at an  
absorbing four-port device of given complex refractive-index profile. 
In this paper we apply these results to study
the entanglement properties influenced by 
propagation in real dielectrics and
extend the theory also to amplifying four-port devices.
Enlarging the system by introducing appropriately
chosen auxiliary degrees of freedom, we first construct the unitary
transformation in the enlarged Hilbert space. Taking the  
trace with regard to the auxiliary variables, we then obtain
the sought formulas for the transformation of arbitrary 
input quantum states. Finally, we discuss some applications, 
with special emphasis on the dependence of entanglement
on absorption and amplification.

The paper is organized as follows. In Sec.~\ref{basic} the basic
equations are reviewed and the general transformation formulas
are derived. Examples of possible applications are discussed in
Sec.~\ref{Applications}, and some conclusions are given in 
Sec.~\ref{Conclusion}. 


\section{Quantum state transformations}
\label{basic}


\subsection{Basic equations}

\begin{figure}[h]
\hspace{1cm}
\psfig{file=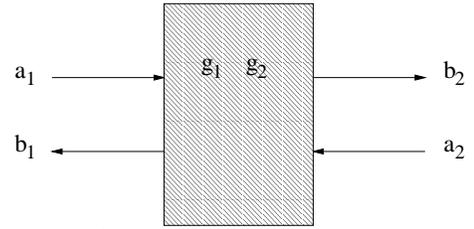,width=6cm}
\caption{\label{device} Quasi one-dimensional geometry of the device
with definition of field input and output operators.}
\end{figure}

Let us briefly review some basic formulas needed for the following
calculations. For simplicity, we restrict ourselves to a
quasi-one-dimensional scheme (Fig.~\ref{device}). The action of the
dielectric device on the incoming radiation is described by means of
the characteristic $2$ $\!\times$ $\!2$ transformation and absorption
matrices ${\bf T}(\omega)$ and ${\bf A}(\omega)$ respectively, which 
are derived in \cite{Gruner96b} on the basis of the quantization scheme 
in \cite{Gruner96a}. They are given in terms of the complex 
refractive-index profile $n(x,\omega)$ of the device. Let 
$\hat{a}_i(\omega)$ and $\hat{b}_i(\omega)$, $i$ $\!=$ $\!1,2$, 
be the amplitude operators of the incoming and outgoing
damped waves at frequency $\omega$. Taking their spatial arguments
at the boundary of the device, we may regard them as being
effectively bosonic operators \cite{Gruner96b}. Further, let
$\hat{g}_i(\omega)$ be the bosonic operators of the 
device excitations, which play the role of operator noise forces
associated with absorption or amplification.
%
%
Introducing the 
two-vector notation $\hat{\bf a}(\omega)$, $\hat{\bf b}(\omega)$ and
$\hat{\bf g}(\omega)$, for the field and device operators
respectively, we may write the input-output relation for 
radiation at an absorbing or amplifying device in the compact form
\begin{equation}
\label{2.4}
\hat{\bf b}(\omega) = {\bf T}(\omega) \hat{\bf a}(\omega) 
+ {\bf A}(\omega) \hat{\bf d}(\omega) ,
\end{equation}
where the transformation and absorption matrices satisfy the relation 
\begin{equation}
\label{2.5}
{\bf T}(\omega) {\bf T}^+(\omega) +\sigma {\bf A}(\omega) {\bf
A}^+(\omega) =  {\bf I}, 
\end{equation}
and $\sigma$ $\!=$ $\!+1$, $\hat{\bf d}(\omega)$ 
$\!=$ $\!\hat{\bf g}(\omega)$ for absorbing devices and 
\mbox{$\sigma$ $\!=$ $\!-1$}, $\hat{\bf d}(\omega)$ $\!=$ 
$\!\hat{\bf g}^\dagger(\omega)$ for amplifying devices. 
The above given equations are valid for any chosen frequency. Knowing 
the amplitude operators as functions of frequency, the full-field
operators can be constructed by appropriate integration over
the frequency in a straightforward way \cite{Gruner96b}.


\subsection{Unitary operator transformations}
\label{uniop}

The operator input-output relation (\ref{2.4}) contains all the 
information necessary to transform an arbitrary function of the 
input-field operators into the corresponding function
of the output-field operators. 
In particular, it enables one to express arbitrary
moments and correlations of the outgoing field
in terms of those of the incoming field and the
device excitations, and hence all knowable information
about the quantum state of the outgoing field can be
obtained. Commonly, quantum states are expressed in
terms of density matrices or phase-space functions  
 -- representations that are more suited to study a 
quantum state as a whole. 

In order to calculate the density operator of the
outgoing field for both absorbing and amplifying devices, we 
follow the line given in \cite{Knoll99} for absorbing devices.
We first define the four-vector operators
\begin{equation}
\label{3.1}
\hat{\mbb{\alpha}}(\omega) = \left( \begin{array}{c}
\hat{\bf a}(\omega) \\ \hat{\bf d}(\omega) \end{array} \right) \,,\quad
\hat{\mbb{\beta}}(\omega) = \left( \begin{array}{c}
\hat{\bf b}(\omega) \\ \hat{\bf f}(\omega) \end{array} \right),
\end{equation}
where $\hat{\bf f}(\omega)$ $\!=$ $\!\hat{\bf h}(\omega)$
for an absorbing device, and $\hat{\bf f}(\omega)$ $\!=$ 
$\!\hat{\bf h}^\dagger(\omega)$ for an amplifying device,
with $\hat{\bf h}(\omega)$ being some auxiliary bosonic 
(two-vector) operator. The input-output relation (\ref{2.5}) 
can then be extended to the four-dimensional transformation
\begin{equation}
\label{3.2}
\hat{\mbb{\beta}}(\omega) = \mbb{\Lambda}(\omega)
\hat{\mbb{\alpha}}(\omega) 
\end{equation}
with
\begin{equation}
\label{3.3}
\mbb{\Lambda}(\omega) \mbb{J} \mbb{\Lambda}^+(\omega) = \mbb{J},
\qquad \mbb{J} = \left( \begin{array}{cc} {\bf I} & {\bf 0} \\ {\bf 0} 
& \sigma {\bf I} \end{array} \right) .
\end{equation}
Hence, $\mbb{\Lambda}(\omega)$ $\!\in$ SU($4$) for absorbing devices
\cite{Knoll99}, and $\mbb{\Lambda}(\omega)$ $\!\in$ SU($2,2$) for
amplifying devices (if an overall phase factor is included
in the input operators). Note, that lossless devices, where 
${\bf A}(\omega)$ $\!\equiv$ $\!{\bf 0}$, can be 
described by SU($2$) group transformations \cite{Campos89,Leonhardt93}.
Since the group SU($4$) is compact, while SU($2,2$) is noncompact,
qualitatively different properties of the state transformations
are expected to occur in these two cases. Introducing
the (commuting) positive Hermitian matrices
\begin{equation}
\label{3.4}
{\bf C}(\omega) = \sqrt{{\bf T}(\omega){\bf T}^+(\omega)},
\quad 
{\bf S}(\omega) = \sqrt{{\bf A}(\omega){\bf A}^+(\omega)} ,
\end{equation}
which, by Eq.~(\ref{2.5}), obey the relation ${\bf C}^2(\omega)$ $\!+$
$\!\sigma {\bf S}^2(\omega)$ $\!=$ $\!{\bf I}$,   
it is not difficult to generalize
the matrix $\mbb{\Lambda}(\omega)$ in \cite{Knoll99} to
\begin{equation}
\label{3.5}
\mbb{\Lambda}(\omega) = \left( \begin{array}{cc} {\bf T}(\omega) &
{\bf A}(\omega) \\ -\sigma {\bf S}(\omega) {\bf C}^{-1}(\omega) {\bf
T}(\omega) & {\bf C}(\omega) {\bf S}^{-1}(\omega) {\bf A}(\omega)
\end{array} \right) .
\end{equation}

Both the SU($4$) and SU($2,2$) group elements can be written in
exponential form 
\begin{equation}
\label{3.6}
\mbb{\Lambda}(\omega) = {\rm e}^{-i\mbb{\Phi}(\omega)} \,, \quad
\mbb{\Phi}^+(\omega) = \mbb{J} \mbb{\Phi}(\omega) \mbb{J} ,
\end{equation}
and a unitary operator transformation
\begin{equation}
\label{3.7}
\hat{\mbb{\beta}}(\omega) = \hat{U}^\dagger \hat{\mbb{\alpha}}(\omega) 
\hat{U}
\end{equation}
can be constructed, where
\begin{equation}
\label{3.8}
\hat{U} = {\rm exp} \Bigg\{ -i \int_0^\infty d\omega \left[
\hat{\mbb{\alpha}}^\dagger(\omega) \right]^T \mbb{J}
\mbb{\Phi}(\omega) \hat{\mbb{\alpha}}(\omega) \Bigg\} .
\end{equation}
Note that the unitarity of $\hat{U}$ follows directly from Eq.~(\ref{3.6}). 

Let the density operator of the input quantum state be a
functional of $\hat{\mbb{\alpha}}(\omega)$ and
$\hat{\mbb{\alpha}}^\dagger(\omega)$, $\hat{\varrho}_{\rm in}$ $\!=$
$\!\hat{\varrho}_{\rm in}[\hat{\mbb{\alpha}}(\omega),
\hat{\mbb{\alpha}}^\dagger(\omega)]$. The density operator of 
the quantum state of the outgoing fields can then be given by
\begin{eqnarray}
\label{3.9}
\lefteqn{
\hat{\varrho}^{(F)}_{\rm out} = {\rm Tr}^{(D)} \left\{ \hat{U}
\hat{\varrho}_{\rm in} \hat{U}^\dagger \right\}
}
\nonumber \\ && 
= {\rm Tr}^{(D)} \left\{
\hat{\varrho}_{\rm in} \left[ \mbb{J} \mbb{\Lambda}^+(\omega) \mbb{J}
\hat{\mbb{\alpha}}(\omega), \mbb{J} \mbb{\Lambda}^T(\omega) \mbb{J}
\hat{\mbb{\alpha}}^\dagger(\omega) \right] \right\} ,
\end{eqnarray}
where ${\rm Tr}^{(D)}$ means trace with respect to the device variables.
It should be pointed out that $\hat{\varrho}^{(F)}_{\rm out}$ 
in Eq.\,(\ref{3.9}) does not depend on the auxiliary variables 
introduced in Eq.~(\ref{3.2}).
The SU(4)-group transformation preserves operator ordering and thus
for absorbing devices, the $s$-parametrized phase-space functions 
transform as
\begin{equation}
\label{3.9a}
P_{\rm out}[\mbb{\alpha}(\omega);s] =
P_{\rm in}[\mbb{\Lambda}^+(\omega)\mbb{\alpha}(\omega);s].
\end{equation}
Since the SU(2,2)-group transformation mixes creation and 
annihilation operators, an equation of the type (\ref{3.9a}) is not
valid for amplifying devices in general. An exception is
the Wigner function that corresponds to symmetrical 
ordering ($s$ $\!=$ $\!0$):
\begin{equation}
\label{3.10}
W_{\rm out}\left[\mbb{\alpha}(\omega)\right] 
= W_{\rm in}[\mbb{J} \mbb{\Lambda}^+(\omega) \mbb{J} \mbb{\alpha}(\omega)] .
\end{equation}
For amplifying devices, the calculation of the output state is 
rather involved in general. Formulas for Fock-state transformation
are given in the Appendix. 


\section{Applications}
\label{Applications}

As already mentioned, the input-output relation (\ref{2.4})
enables one to calculate arbitrary moments and correlations
of the outgoing field. It is worth noting that there is
no need to introduce fictitious beam splitters for modeling 
the losses. The transmittance and absorption
matrices in Eq.~(\ref{2.4}) automatically take account
of the losses, because they are calculated from Maxwell's
equations with complex permittivity. To give an example,
we compute in Sec.~\ref{visibility} the visibility of
interference fringes in photon-number detection in a 
Mach-Zehnder interferometer with lossy beam splitters.

The input-output relation (\ref{3.9}) can advantageously
be used when knowledge of the transformed quantum state as a whole is
required. This is typically the case in quantum communication,
which is essentially based on entangled quantum states. 
For quantification of entanglement -- a quantum-coherence 
property that sensitively responds to losses -- 
information about the full quantum state is needed in general. 
The entanglement measure we use is the quantum relative entropy
(the quantum analog of the classical Kullback--Leibler entropy)
defined by \cite{Vedral98}
\begin{equation}
\label{mix1}
E(\hat{\sigma})= \min_{\hat{\rho}\in{\cal S}} {\rm Tr}
\left[ \hat{\sigma} \left( \ln \hat{\sigma} -\ln \hat{\rho} \right)
\right], 
\end{equation}
with $\hat{\sigma}$ and ${\cal S}$ being, respectively, the bipartite 
quantum state under study and the set of all separable quantum states.
We stress here that the relative entropy is indeed a ``good''
entanglement measure, because it satisfies the necessary conditions 
that should be required of such a measure 
\cite{Vedral98,Vedral97,Horodecki99}. 
Note that any proper entanglement measure satisfying them
would do (the Bures metric being another typical example).

In Sec.~\ref{mix} we study the entanglement produced at a
realistic beam splitter by initially uncorrelated photons, 
and in Sec.~\ref{trans} we analyze the degradation of  
entanglement during propagation through lossy media,
with special emphasis on Bell-type states. 
Effects associated with amplification are addressed in 
Sec.~\ref{amplification}.


\subsection{Visibility of interference fringes}
\label{visibility}

\begin{figure}[ht]
\psfig{file=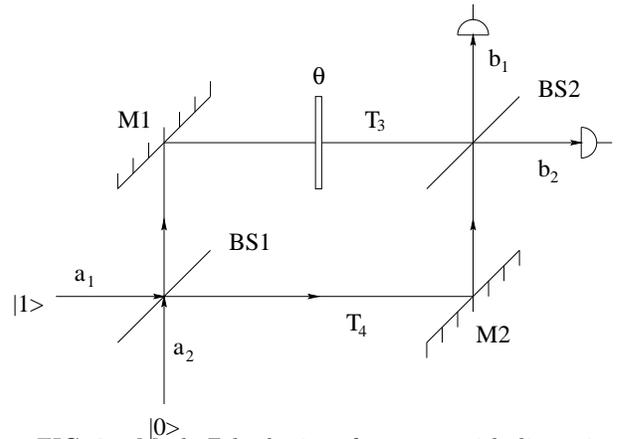,width=8cm}
\caption{\label{fringe} Mach--Zehnder interferometer with dispersive
and absorbing beam splitters BS1 and BS2 and a phase shift
\protect$\Theta$. The mirrors M1 and M2 as well as the branches
between BS1 and BS2 are assumed to be lossy with transmittance $T_3$
(upper branch) and $T_4$ (lower branch), respectively.}
\end{figure}

To give an example of application of the input-output relations
(\ref{2.4}), we consider the visibility of interference fringes in a
Mach--Zehnder interferometer in Fig.~\ref{fringe}. A sin\-gle 
photon is fed into one input port, the other input port being unused.
%
%
The quantity we are interested in is the visibility 
\begin{equation} 
\label{fr1}
V_k = \frac{\langle \hat{n}_k \rangle_{\rm max}
-\langle \hat{n}_k \rangle_{\rm min}}
{\langle \hat{n}_k \rangle_{\rm max}
+\langle \hat{n}_k \rangle_{\rm min}} \,,
\end{equation}
where $\langle \hat{n}_k \rangle_{\rm max}$ 
($\langle \hat{n}_k \rangle_{\rm min}$) is the
maximum (minimum) value of the mean photon number in the
$k$th output channel ($k$ $\!=$ $\!1,2$). 

In order to model the
losses in the interferometer arms
(e.g., nonperfect mirrors or dissipation processes 
in optical fibers connecting the beam splitters BS1 and BS2),
in \cite{Hendrych} a fictitious (nonabsorbing) beam splitter 
is inserted into each branch of the interferometer.
In practice, however, the beam splitters BS1 and BS2 are also
expected to give rise to some losses. Whereas the losses arising 
from the beam splitter BS1 may be thought of as being included
in the replacement scheme considered in \cite{Hendrych}, 
inclusion in the calculation of the losses arising from the beam 
splitter BS1 would require that two additional fictitious  
beam splitters were inserted between the beam splitter BS2 and 
the detectors. Altogether, a replacement scheme with
four fictitious beam splitters at least must be considered in
order to model all the losses.

Application of the input-output relations
(\ref{2.4}) shows that there is no need for such an involved
replacement scheme. Instead, the proper transmittance and
reflection coefficients of the 
beam splitters and mirrors (or fibers)
can be used to obtain the correct physics, including the losses.   
Applying the input-output relations (\ref{2.4}) successively 
to the beam splitter BS2, the lossy branches, and the beam splitter
BS1 and assuming the devices are in the vacuum state, so that the 
overall input state is $|\psi_{\rm in}\rangle$ $\!=$ $\!|1,0,0,0\rangle$,
it is not difficult to show that
\begin{eqnarray}
\langle \hat{n}_1 \rangle &=& |R_1|^2 |T_3|^2 |R_2|^2
+|T_1|^2 |T_4|^2 |T_2|^2 \nonumber \\ &&
+ 2 |R_1||R_2||T_1||T_2||T_3||T_4|
\cos\Theta_1 \,, 
\end{eqnarray}
\begin{eqnarray}
\label{fr0}
\langle \hat{n}_2 \rangle &=& |R_1|^2 |T_3|^2 |T_2|^2
+|R_2|^2 |T_4|^2 |T_1|^2 \nonumber \\ &&
+ 2 |R_1||R_2||T_1||T_2||T_3||T_4|
\cos\Theta_2 \,, 
\end{eqnarray}
where
$\Theta_1$ $\!=$ $\!\Theta$ 
$\!+$ $\!\varphi_{R_1}$ $\!-$ $\!\varphi_{T_1}$
$\!+$ $\!\varphi_{R_2}$ $\!-$ $\!\varphi_{T_2}$
$\!+$ $\!\varphi_{T_3}$ $\!-$ $\!\varphi_{T_4}$
and
$\Theta_2$ $\!=$ $\!\Theta$
$\!+$ $\!\varphi_{R_1}$ $\!-$ $\!\varphi_{T_1}$
$\!-$ $\!\varphi_{R_2}$ $\!+$ $\!\varphi_{T_2}$
$\!+$ $\!\varphi_{T_3}$ $\!-$ $\!\varphi_{T_4}$.
Here and in the following, the notation
$(T_l)_{11}$ $\!=$ $\!(T_l)_{22}$ $\!\equiv$ $\!R_l$ $\!=$ 
$\!|R_l|e^{i\varphi_{R_l}}$
and 
$(T_l)_{12}$ $\!=$ $\!(T_l)_{21}$ $\!\equiv$ $\!T_l$ $\!=$ 
$\!|T_l|e^{i\varphi_{T_l}}$
for the elements of the transmittance matrix 
{\boldmath $T$}$_l$ of the $l$th four-port device
is used [$l$ $\!=$ $\!1,2$, beam splitters BS1 and BS2;
$l$ $\!=$ $\!3,4$, upper (3) and lower (4) branch of the 
interferometer]. Note that for a lossy device 
\begin{equation}
\arg R_l - \arg T_l \neq \pi/2
\label{phase}
\end{equation}
in general.
Combining Eqs.~(\ref{fr1}) -- (\ref{fr0}), we easily derive
\begin{eqnarray}
\label{fr2}
V_1 &=& 2 \left( \frac{|R_1||T_3||R_2|}{|T_1||T_4||T_2|}
+\frac{|T_1||T_4||T_2|}{|R_1||T_3||R_2|} \right)^{-1} ,
\\ \label{fr3}
V_2 &=& 2 \left( \frac{|R_1||T_3||T_2|}{|R_2||T_1||T_4|}
+\frac{|R_2||T_1||T_4|}{|R_1||T_3||T_2|} \right)^{-1}\,.
\end{eqnarray}
It is worth noting that Eqs.\,(\ref{fr2}) and (\ref{fr3})
are valid for the really observed reflection and transmission
coefficients $R_k$ and $T_k$, respectively, with
\begin{equation}
|R_k|^2+|T_k|^2 \leq 1.
\label{3A.6}
\end{equation}
The equality sign would be realized for nonabsorbing devices. 

Comparing with the formulas for the visibilities derived
in \cite{Hendrych}, we observe that they look like Eqs.~(\ref{fr2}) 
and (\ref{fr3}). However, this resemblance is only formal. In fact,
all the reflection and transmission coefficients (including the phases) 
introduced in \cite{Hendrych} satisfy the relations valid for nonabsorbing 
devices and therefore differ from the measured reflection and 
transmission coefficients that in a real experiment determine 
the fringe visibilities. Even if additional fictitious beam 
splitters were included in the model in \cite{Hendrych}, there 
would be no unique relation between auxiliary and
actual parameters in general.


\subsection{Photon entanglement at a beam splitter}
\label{mix}

\begin{figure}[h]
\psfig{file=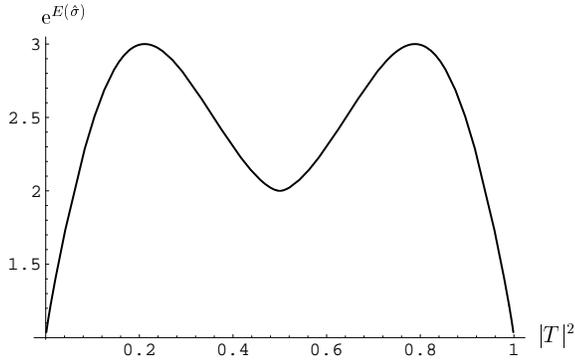,width=8cm}
\caption{\label{lossless} Entanglement created at a lossless beam
splitter with both input modes excited in an one-photon Fock
state. The entanglement is shown in exponential scaling
vs. transmittance. Maximal entanglement \protect$\ln 3$ is created at a
\protect$20\%/80\%$ beam splitter whereas for a symmetric device it
drops to \protect$\ln 2$ due to destructive interference.}
\end{figure}

Superimposing two nonclassically excited modes by a lossless
beam splitter, one can generate entangled states with interesting 
properties \cite{Campos89}. Two of the simplest
examples are as follows. Having a single-photon Fock state in one
input channel of a 50\%-50\% beam splitter, i.e., $|T|^2$ $\!=$ 
$\!|R|^2$ $\!=$ $\!1/2$, whereas the other input channel is unused, 
the output state is a superposition of states with the photon in 
one of the output channels. If each of the two incoming modes is 
prepared in a single-photon Fock state, then the output state is a 
superposition of states with two photons in one output channel.
In either case, the output state is a superposition of two
states and the maximum entanglement of $\ln 2$ (which corresponds to 
$1$ bit) is realized. Note that for pure states the entanglement measure 
(\ref{mix1}) reduces to the von Neumann entropy of one subsystem.
When $|T|^2$ $\!\neq$ $\!|R|^2$ then the 
output state in the latter case is a superposition of three states, 
because each outgoing mode can now contain either zero, one, or two 
photons. The maximum entanglement of a three-state system is $\ln 3$,
which is realized if \mbox{$|T|^2$ $\!=$ $\!1/2$ $\!\times$ $\!(\!1$ $\!\pm$
$\!1/\sqrt{3})$} (see Fig.~\ref{lossless}). 
Hence, a non-50\%-50\% beam splitter can produce
stronger entanglement than a 50\%-50\% beam splitter which
suppresses one possible outcome owing to interference. With regard to 
entanglement, this interference effect is thus destructive. 
%
%

\begin{figure}[h]
\psfig{file=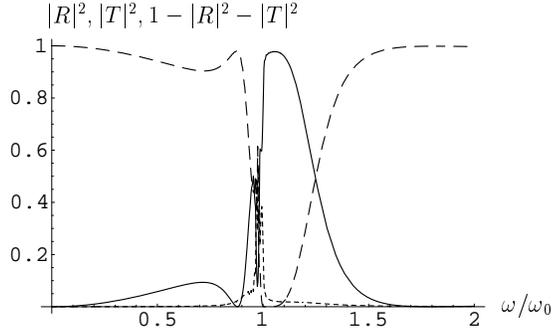,width=8cm}
\caption{\label{slab} The reflection coefficient \protect$|R|^2$
(full line), the transmission coefficient \protect$|T|^2$ (dashed
line), and the absorption coefficient 
\mbox{\protect$(1$ $\!-$ $|R|^2$ $\!-$ $\!|T|^2)$} (dotted line)
of a dielectric plate are shown as 
functions of frequency \protect$\omega$ for
$\epsilon_{\rm s}$ $\!=$ $\!1.5$ and $\gamma/\omega_0$ $\!=$ $\!0.001$
in Eq.~(\protect\ref{mix2}), and the plate thickness $2c/\omega_0$.}
\end{figure}

Let us now raise the question of the amount of entanglement achievable 
in case of a realistic beam splitter -- a question that
may be important for the quality of quantum communication by means 
of entangled photonic states obtained by available devices.
The question can be answered by applying the input-output relation
(\ref{3.9}) and calculating the output state of the interfering
modes obtained by an absorbing beam splitter. To give an example, 
let us study the entanglement produced by a 
dielectric plate of permittivity 
\begin{equation}
\label{mix2}
\epsilon(\omega) = 1+ \frac{\epsilon_s-1}{1-(\omega/\omega_0)^2 -2i\gamma
\omega/\omega_0^2}
\end{equation}
($\epsilon_s$ $\!=$ $\!1.5$)
and thickness $d$ $\!=$ $\!2c/\omega_0$ for the case where
either one or each of the two incoming modes is prepared in a 
single-photon Fock state.
%
%
The squares of the absolute values of the calculated reflection,
transmission, and absorption coefficients as functions of
frequency \cite{Gruner96b} are shown in Fig.~\ref{slab} for
$\gamma$ $\!=$ $\!0.001$. When the device is not excited, then the 
overall input state is either $|1,0,0,0\rangle$ or $|1,1,0,0\rangle$ 
for the two cases under consideration. The resulting mixed states 
of the outgoing modes are calculated in \cite{Knoll99}. 
Here, we have calculated the amount of entanglement of the 
states using the definition (\ref{mix1}). 

\begin{figure}[h]
\psfig{file=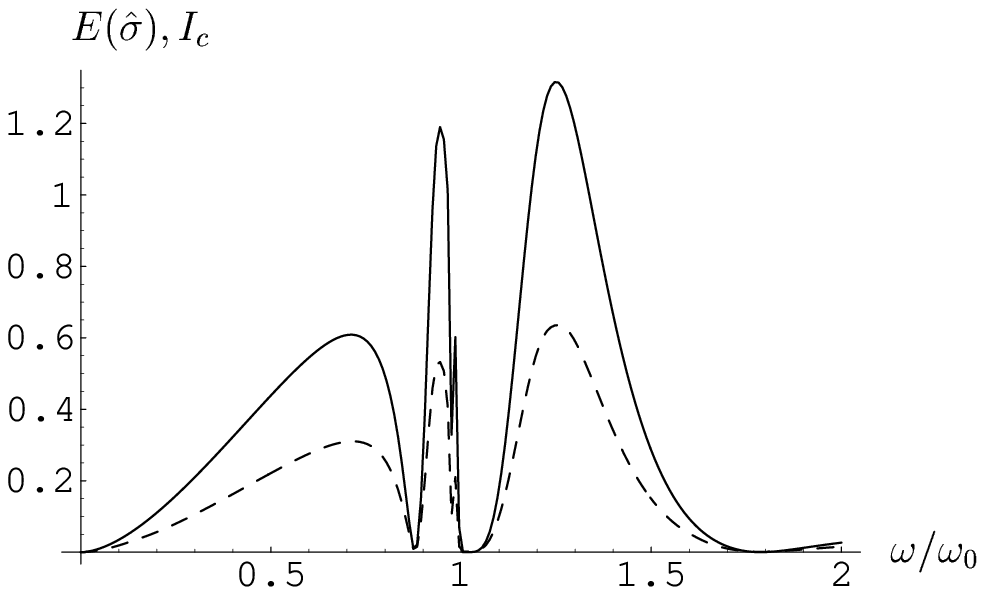,width=8cm}
\caption{\label{creation10} Frequency dependence of the entanglement
measure \protect$E(\hat{\sigma})$ (dashed line) and the total
correlation $I_c$ (solid line) for
a state \protect$|1,0,0,0\rangle$ impinging on a beam splitter with
$\gamma/\omega_0$ $\!=$ $\!0.001$ in Eq.~(\protect\ref{mix2}) and the
other parameters given in the text.} 
\end{figure}

\begin{figure}[h]
\psfig{file=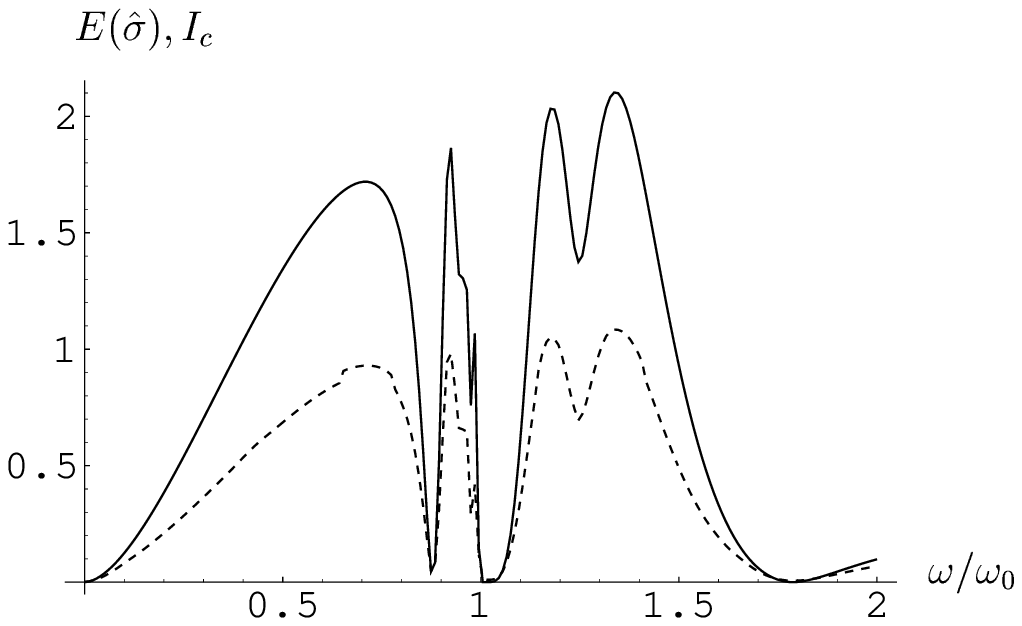,width=8cm}
\caption{\label{creation11} Frequency dependence of the entanglement
measure \protect$E(\hat{\sigma})$ (dashed line) and the total
correlation $I_c$ (solid line) for
a state \protect$|1,1,0,0\rangle$ impinging on a beam splitter with
$\gamma/\omega_0$ $\!=$ $\!0.001$ in Eq.~(\protect\ref{mix2}) and the
other parameters given in the text.} 
\end{figure}

\begin{figure}[h]
\psfig{file=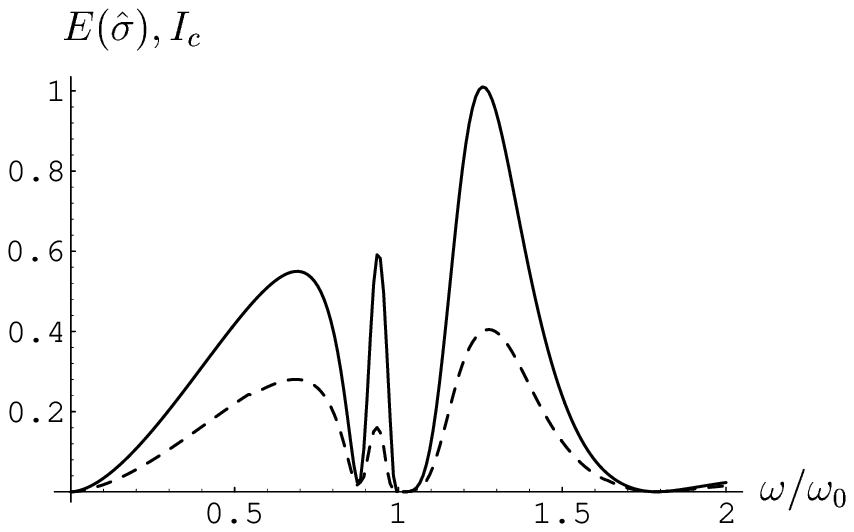,width=8cm}
\caption{\label{gamma01} Frequency dependence of the entanglement
measure \protect$E(\hat{\sigma})$ (dashed line) and the total
correlation $I_c$ (solid line) for a state \protect$|1,0,0,0\rangle$
impinging on a beam splitter with \protect$\gamma/\omega_0$ $\!=$
$\!0.01$ in Eq.~(\protect\ref{mix2}) and the other parameters given in
the text.}
\end{figure}

Results are plotted in 
Figs.~\ref{creation10} and \ref{creation11} for $\gamma$ $\!=$ $\!0.001$, 
and in Fig.~\ref{gamma01} for \mbox{$\gamma$ $\!=$ $\!0.01$}. 
For comparison, the figures also show the mutual information 
\mbox{$I_c$ $\!=$ $\!S_1$ $\!+$ $\!S_2$ $\!-$ $\!S_{12}$}, 
where $S_1$ and $S_2$ are the 
von Neumann entropies of the outgoing modes $1$ and $2$,
respectively, and $S_{12}$ is the entropy of the composite 
two-mode system. Obviously, the mutual information 
may be regarded as a measure of the total amount of
correlation contained in the states. In regions where the 
absorption is sufficiently weak, the output 
state is almost pure, and thus \mbox{$S_{12}$ $\!\approx$
$\!0$} and \mbox{$E(\hat\sigma)$ $\!\approx$ $\!S_i$}. 
Hence, the two curves in Figs.~\ref{creation10} and 
\ref{creation11} 
differ there only by a factor approximately equal to two.
%
%
With increasing absorption the two curves cannot be related
to each other by simple scaling, as it can be seen from
Fig.~\ref{gamma01}. In particular, the maximally achievable 
amount of entanglement of about $0.4$ is much less than  
$\ln 2$ achievable with a lossless device. 
 
{F}rom Figs.~\ref{creation10} and \ref{creation11} strong reduction
of entanglement is observed in the resonance region. Here reflection
and absorption are strongest, so that the two modes are only weakly
mixed and absorption prevents the device from creating quantum coherence.  
As expected, substantial entanglement is observed in regions where
the absorption is weak and $|T|^2$ and $|R|^2$ nearly satisfy the
condition of maximum entanglement. In Fig.~\ref{creation10}
this is the case for $\omega/\omega_0$ $\!\approx$ $\!1.25$ where 
the value of entanglement becomes close to the maximally achievable 
value of $\ln 2$. In Fig.~\ref{creation11} the value
of entanglement becomes close to the maximally achievable
value of $\ln 3$ at $\omega/\omega_0$ $\!\approx$ $\!1.18$
and $\omega/\omega_0$ $\!\approx$ $\!1.33$.
The relative minimum in Fig.~\ref{creation11} at 
$\omega/\omega_0$ $\!\approx$ $\!1.25$
indicates the effect of destructive interference mentioned above. 

The results show that entanglement sensitively depends 
on the optical properties of the material used for manufacturing the
optical device. They demonstrate the importance of optimizing the
frequency regime of quantum communication schemes with given devices.


\subsection{Entangled-state transmission through a lossy channel}
\label{trans}

\subsubsection{Bell-type basis states \protect$|\Psi_n^\pm\rangle$}

Let us now turn to the question of entanglement degradation
during the propagation through dielectric matter such as
an optical fiber. For this purpose, we consider two modes each of 
which propagates through a dielectric medium of complex
permittivity. Assuming the incoming 
modes are prepared in a maximally entangled Bell-type state
\begin{equation}
\label{trans0}
|\Psi_n^\pm\rangle = \frac{1}{\sqrt{2}}
\left( |0n\rangle \pm |n0 \rangle \right), 
\end{equation}
we apply Eq.~(\ref{3.9}) and calculate the
quantum state of the two outgoing modes. After some
algebra we derive 
\begin{eqnarray}
\label{trans3}
\hat{\varrho}_{\rm out}^{({\rm F})} &=& \frac{1}{2} \Bigg[
\sum\limits_{k=0}^{n-1} {n \choose k} |T_1|^{2k}
\left( 1-|T_1|^2 \right)^{n-k} |k0\rangle\langle k0|
\nonumber \\ && \hspace{2ex}
+\sum\limits_{k=0}^{n-1} {n \choose k} |T_2|^{2k}
\left( 1-|T_2|^2 \right)^{n-k} |0k\rangle\langle 0k| \Bigg]
\nonumber \\ && \hspace{2ex}
+{\textstyle\frac{1}{2}}\left(|T_1|^{2n}+|T_2|^{2n}\right)
\,|\Psi'_n{^{\!\pm}} \rangle\langle \Psi'_n{^{\!\pm}}|,
\end{eqnarray}
where
\begin{equation}
\label{trans3a}
|\Psi'_n{^{\!\pm}} \rangle = \left(|T_1|^{2n}\!+\!|T_2|^{2n}\right)^{-1/2}
\big( T_1^n |n0\rangle \pm T_2^n |0n\rangle \big).
\end{equation}
Note that when setting $n$ $\!=$ $\!1$, the transformation 
of the ordinary Bell basis states $|\Psi^\pm\rangle$ 
$\!\equiv$ $\!|\Psi_1^\pm\rangle$ are obtained. 
In what follows we assume that the transmission coefficients
$T_k$ \mbox{($k$ $\!=$ $\!1,2$)} are given by 
\begin{equation}
\label{trans8}
T_k = T_k(\omega) = {\rm e}^{in_k(\omega)\omega l_k/c} ,
\end{equation}
with $n_k(\omega)$ $\!=$ $\!\sqrt{\epsilon_k(\omega)}$ $\!=$
$\eta_k(\omega)+i\kappa_k(\omega)$ and $l_k$ being
the complex refractive indexes of the media and the 
propagation lengths, respectively. According to the 
Lambert--Beer law, $|T_k|$ decreases exponentially with the 
length of propagation: $|T_k|$ $\!=$ $\!\exp(-l_k/L_k)$, 
$L_k$ $\!=$ $\!c/(\omega\kappa_k)$.
In special cases when one mode propagates through vacuum,
$n(\omega)$ $\!=$ $\!1$, the corresponding transmission coefficient,
by Eq.~(\ref{trans8}), is just a phase factor. 

For a first insight into the behavior of the transmitted quantum 
state it may be instructive to look at the overlap 
of the output state with the input state, which is 
\begin{equation}
\label{trans1b}
\langle\Psi_n^\pm|\hat{\varrho}_{\rm out}^{({\rm F})}|\Psi_n^\pm\rangle 
= {\textstyle\frac{1}{4}} 
\left( |T_1|^{2n}\!+\!|T_2|^{2n} \!+\! T_1^{\ast n} T_2^n 
\!+ \!T_1^n T_2^{\ast n}\right).
\end{equation}
We see that the characteristic length of degradation of the overlap 
(fidelity) is not given by $L_k$ but by the shorter length $L_k/(2n)$.  
Hence, the overlap rapidly approaches zero with increasing number 
of photons even for weak damping of the intensity or related
(classical) quantities.
 
As already mentioned, a proper measure of entanglement is the quantum 
relative entropy defined by Eq.~(\ref{mix1}). In order to estimate an 
upper bound, we employ the convexity property \cite{Wehrl78}
\begin{equation}
\label{konvex}
E[\lambda\hat{\sigma}_1 +(1-\lambda) \hat{\sigma}_2] \le
\lambda E(\hat{\sigma}_1) +(1-\lambda) E(\hat{\sigma}_2) \,.
\end{equation}
{F}rom Eq.~(\ref{trans3}) it is seen that 
$\hat{\varrho}_{\rm out}^{(F)}$ has the form
\begin{equation}
\label{konvex1}
\hat{\varrho}_{\rm out}^{(F)} 
= \lambda\hat\sigma_1 + (1 - \lambda) \hat\sigma_2,
\end{equation}
where $\hat\sigma_1$ is a separable state 
[$E(\hat\sigma_1)$ $\!=$ $\!0$] and 
\begin{equation}
\label{konvex2}
\hat\sigma_2 = |\Psi'_n{^{\!\pm}}\rangle \langle \Psi'_n{^{\!\pm}}|
\end{equation}
is a pure state, the entanglement of which is simply given by 
the entropy of one of the two modes. We thus find
\begin{equation}
\label{ent2a}
E(\hat{\varrho}_{\rm out}^{({\rm F})}) \leq B,
\end{equation}
\begin{eqnarray}
\label{ent2}
\lefteqn{
B = \frac{1}{2} \big[
\left( |T_1|^{2n} +|T_2|^{2n} \right) \ln\! \left( |T_1|^{2n} +
|T_2|^{2n} \right)
} 
\nonumber \\ && \hspace{7ex} 
-\,|T_1|^{2n} \ln |T_1|^{2n} -|T_2|^{2n}
\ln |T_2|^{2n} \big] \,. 
\end{eqnarray}
In particular when $T_1$ $\!=$ $\!T_2$ $\!=$ $T$, then
\begin{equation}
\label{ent3}
E(\hat{\varrho}_{\rm out}^{({\rm F})}) \le |T|^{2n} \ln 2 
= {\rm e}^{- 2nl/L} \ln 2 ,
\end{equation}
i.e., the characteristic length of entanglement degradation
decreases as $1/(2n)$ at least. 
The result (\ref{ent3}) reveals that
with an increasing number of photons the quantum interference 
relevant for entanglement exponentially decreases at least. 
Such a behavior is typical of quantum decoherence phenomena 
and is not restricted to Fock states. 

It should be mentioned that for a pair of spin-$\frac{1}{2}$ 
parties a decomposition of the density matrix into a separable 
part and a single pure state is always possible \cite{Lewenstein98}.
Moreover, there exists a unique maximal
$\lambda$ such that the inequality (\ref{konvex}) reduces to an 
equality and thus $(1$ $\!-$ $\!\lambda)E(\hat{\sigma}_2)$ becomes 
a measure of entanglement. However, for larger dimensions of the 
Hilbert space we are left with the general inequality (\ref{konvex}).

\begin{figure}[h]
\psfig{file=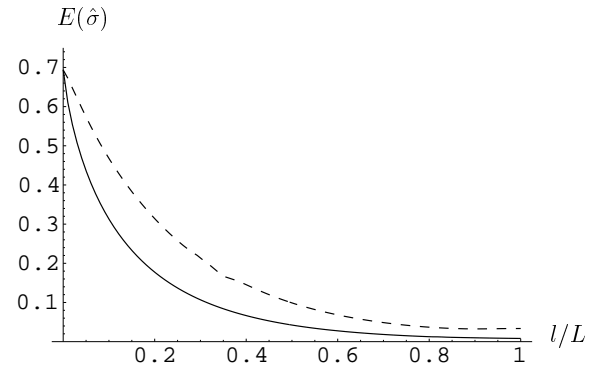,width=8cm}
\caption{\label{zweifaser} Entanglement degradation of a singlet state
\protect$|\Psi^-_n\rangle\langle \Psi^-_n|$
[Eq.~(\protect\ref{trans0})]
with one photon (\protect$n\!=\!1$ full curve) and two photons
(\protect$n\!=\!2$ dashed curve) after transmission 
through absorbing channels of equal transmittance.}
\end{figure}

Examples of entanglement degradation [calculated on the basis 
of Eq.~(\ref{mix1})] for singlet states with one photon,
$|\Psi_1^{\pm}\rangle$, and two photons, $|\Psi_2^{\pm}\rangle$,
are shown in Fig.~\ref{zweifaser} for the case where
the two modes propagate in equal media over equal distances. 
%
%
We observe that for the state $|\Psi_2^{\pm}\rangle$ the 
upper bound ${\rm e}^{-4l/L} \ln 2$ defined by the inequality
(\ref{ent3}) is a very good approximation to the entanglement at
propagation length $l$. In contrast, for the state
$|\Psi_1^{\pm}\rangle$ the actual values of entanglement are
typically smaller than it might be expected from the upper bound ${\rm
e}^{-2l/L} \ln 2$. 
Since for $n$ $\!>$ $\!2$ the upper bound ${\rm e}^{-2nl/L} \ln 2$
is always smaller than the entanglement observed for the
state $|\Psi_2^{\pm}\rangle$ (at least for \mbox{$0$ $\!<$ $l$ $\!\leq$ 
$\!L$)}, we leave with the result that the two-photon singlet state
$|\Psi_2^{\pm}\rangle$ is the most robust one within
the class of states $|\Psi_n^{\pm}\rangle$. 


\subsubsection{Bell-type basis states \protect$|\Phi_n^\pm\rangle$}

The Bell-type states
\begin{equation}
\label{trans0a}
|\Phi_n^\pm\rangle = \frac{1}{\sqrt{2}}
\left( |00\rangle \pm |nn\rangle \right) 
\end{equation}
can be obtained from the somewhat more general class of states
\begin{equation}
\label{sqvac1}
|\Phi_n^q\rangle = \frac{1}{\sqrt{1+|q|^2}} \big( |00\rangle + q
\, |nn\rangle \big) 
\end{equation}
for $q$ $\!=$ $\!\pm 1$. Obviously, for $n$ $\!=$ $\!1$ and 
small values of $q$ the state $|\Phi_n^q\rangle$
approximates a two-mode squeezed vacuum 
\begin{equation}
\label{sqvac2}
\exp\!\big[\zeta\big(\hat{a}_1\hat{a}_2-\hat{a}_1^\dagger
\hat{a}_2^\dagger\big)\big] |00\rangle 
= \sqrt{1-|q|^2} \, \sum\limits_m q^m |mm\rangle 
\end{equation}
($q$ $\!=$ $\!\tanh \zeta$, $\zeta$ real) used in 
quantum teleportation with
continuous variables \cite{Kimble}. It is not difficult to prove
that the entanglement of $|\Phi_n^q\rangle$ is
\begin{equation}
\label{sqvac3}
E(|\Phi_n^q\rangle\langle\Phi_n^q|) 
= \ln \,(1+|q|^2) -\frac{|q|^2}{1+|q|^2} \ln |q|^2 ,
\end{equation}
which for $|q|$ $\!=$ $\!1$ attains the maximum value of $\ln 2$.

Let again consider two modes propagating through dielectric matter
and assume that the incoming modes are now prepared in a state 
$|\Phi_n^q\rangle$. We again apply Eq.~(\ref{3.9}) and calculate the
quantum state of two modes. The result reads
\begin{eqnarray}
\label{sqvac4}
\lefteqn{
\hat{\varrho}_{\rm out}^{({\rm F})} =
\frac{|q|^2}{1+|q|^2} \Bigg[
\sum\limits_{k_1,k_2=0}^n {n \choose k_1} {n \choose k_2} |T_1|^{2k_1} 
|T_2|^{2k_2}
}
\nonumber \\ &&\hspace{2ex} \times\, 
\left( 1-|T_1|^2\right)^{n-k_1} \left( 1-|T_2|^2\right)^{n-k_2}
|k_1 k_2\rangle\langle k_1 k_2|
\nonumber \\ &&\hspace{2ex}           
-\,|T_1|^{2n} |T_2|^{2n} |nn\rangle\langle nn|
\Bigg] +\frac{1}{1+|q|^2}
\nonumber \\ &&\hspace{2ex} \times\,
\big[ |00\rangle +qT_1^nT_2^n |nn\rangle\big]
\big[ \langle 00| +\left( q T_1^n T_2^n \right)^\ast \langle nn| \big] .
\end{eqnarray}
Again, from the convexity argument, Eq.~(\ref{konvex}),
an upper bound of the entanglement can be derived
\begin{equation}
\label{sqvac5}
B =
\frac{1}{1+|q|^2} 
\left[ (1+|q'|^2 ) \ln (1+|q'|^2 ) -|q'|^2 \ln
|q'|^2 \right] 
\end{equation}
($q'$ $\!=$ $\!qT_1^nT_2^n$). In particular, for 
small values of $q'$ we find by expansion that 
\begin{equation}
\label{sqvac6}
E(\hat{\varrho}_{\rm out}^{({\rm F})}) \lesssim \frac{|q'|^2}{1+|q|^2} 
\left( 1- \ln |q'|^2 \right) +{\cal O}(|q'|^4),
\end{equation}
which shows that the entanglement decreases as
$|q'|^2$ $\!=$ $\!|q|^2|T_1|^{2n}|T_2|^{2n}$. 

\begin{figure}[h]
\psfig{file=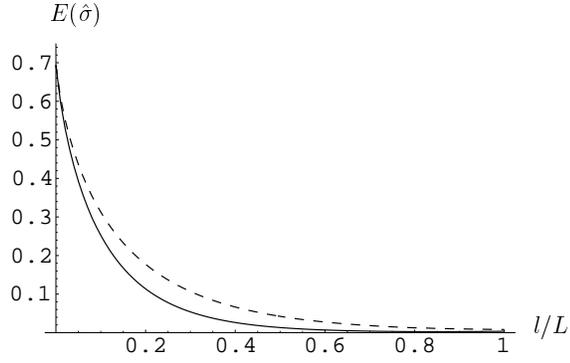,width=8cm}
\caption{\label{vergleich} Comparison of entanglement degradation 
of one-photon Bell basis states 
\protect$|\Phi^\pm\rangle$ (full curve) and \protect$|\Psi^\pm\rangle$ 
(dashed curve).}
\end{figure}

It is also instructive to compare the entanglement degradation of
the states $|\Phi_n^\pm\rangle$ with that of the states 
$|\Psi_n^\pm\rangle$. Similar to the states $|\Psi_n^\pm\rangle$,
within the class of states $|\Phi_n^{\pm}\rangle$ the state 
$|\Phi_2^{\pm}\rangle$ is most robust against entanglement 
degradation. Obviously, the probability of finding $n$
photons in one channel decreases as $|T_i|^n$ for the states 
$|\Psi_n^\pm\rangle$ but decreases as $|T_1T_2|^n$ for 
the states $|\Phi_n^\pm\rangle$.
The entanglement degradation of the states $|\Psi_n^\pm\rangle$
is therefore expected to be less than that of the states 
$|\Phi_n^\pm\rangle$. From Eqs.~(\ref{ent2}) and 
(\ref{sqvac5}) it follows that
\mbox{($|T_1|$ $\!=$ $\!|T_2|$ $\!=$ $\!|T|$ $\!\ll$ $\!1$)} 
\begin{equation}
\label{sqvac7}
\frac{B(|\Phi^\pm\rangle)}{B(|\Psi^\pm\rangle)}
\,\approx\, \frac{|T|^{2n} \left( 1-\ln |T|^{4n} \right)}{2\ln2} \,.
\end{equation}
The numerical results (see Fig.~\ref{vergleich}) indeed
show that the states $|\Psi_n^\pm\rangle$ are more robust
against entanglement degradation that the states $|\Phi_n^\pm\rangle$.
%
%


\subsubsection{Medium with EIT characteristics}

Media having a electromagnetically induced transparency dispersion
characteristics have been of increasing interest \cite{Harris99,Kash99}.
They may offer the possibility of realizing optical quantum gates,
because the group velocity reduction is extremely large such that
there will be plenty of time to manipulate a quantum state
intermediately stored in the medium \cite{Lukin00}. 
The susceptibility of such a medium can be given by
\begin{equation}
\label{trans10}
\chi(\delta) = \frac{N \gamma_1(i\gamma_0 -\delta)}{\Omega^2
+\gamma_\perp\gamma_0 -\delta(\Delta-\delta) + 
i[\delta(\gamma_\perp+\gamma_0)+\Delta\gamma_0]}\,,
\end{equation}
with $\Omega$ being the Rabi frequency if the driving field,
$\gamma_\perp$ the transverse relaxation rate of the probe transition,
$\Delta$ the one-photon detuning, $\gamma_1$ the radiation relaxation
rate of the probe transition, $\gamma_0$ the decay rate of the
ground-state coherence, and $\delta$ the two-photon detuning
(for details, see\cite{Kash99,Lukin00}).

\begin{figure}[h]
\psfig{file=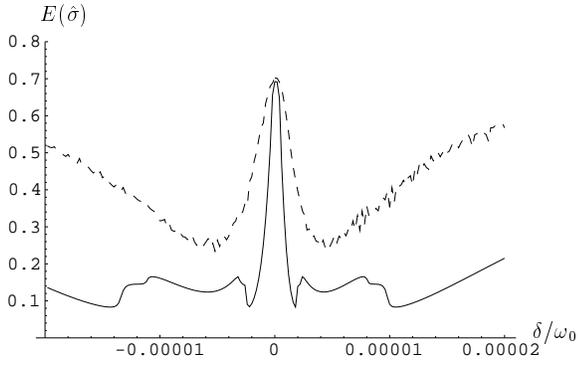,width=8cm}
\caption{\label{eit} Entanglement degradation of a singlet state 
\protect$|\Psi^-\rangle\langle\Psi^-|$ [Eq.~(\protect\ref{trans0})
with \protect$n$ $\!=$ $\!1$]
after transmission of one subsystem (dashed curve) or both subsystems (full
curve) through a medium with susceptibility (\protect\ref{trans10}).}
\end{figure}

We have calculated the degradation of entanglement for the
case where two modes that are initially prepared in a Bell-type state 
$|\Psi_n^\pm\rangle$, Eq.~(\ref{trans0}), can propagate 
through media of that type. Figure \ref{eit} shows
results obtained for ordinary Bell states $|\Psi^{\pm}\rangle$.
In the figure, the two-photon detuning is varied in a small frequency 
region around some optical frequency $\omega_0$.
%
%
The two-peak structure of the absorption coefficient 
[imaginary part of the square root of the susceptibility 
(\ref{trans10})] essentially determines the amount of entanglement
that can be transmitted. It is seen that the initial
entanglement of $\ln 2$ is (approximately) preserved for 
zero two-photon detuning, and the degradation of entanglement 
is almost abrupt for nonzero two-photon detuning. Hence, 
control of entanglement requires fine tuning.


\subsection{Entanglement transformation at amplifying devices}
\label{amplification}

{F}rom Sec.~\ref{basic} we know that quantum-state transformation 
at amplifying four-port devices is connected with SU(2,2) group
transformations. For each frequency, the transformation corresponds 
to the action of a four-mode squeezing operator,
where the destruction (creation) operators of the field modes
are mixed with the creation (destruction) operators of the
device excitations.
Tracing with regard to the device variables then yields the
(two-mode) output state of the field, as is shown
in the Appendix for the case where the (two-mode)
input state of the field is a Fock state and the device 
is in the ground state. 

If the input field is prepared in an entangled state, 
amplification is expected to destroy the entanglement.
Although all necessary formulas are available, the calculation of
the quantum relative entropy is an effort. The number of 
the (real) parameters specifying an arbitrary separable density 
matrix increases dramatically with the dimension of the Hilbert 
space of the subsystems involved. In fact, it is easy to see 
that this number is $[4N^4(N-1)+N^4-1]$, with $N$ being the
Hilbert-space dimension of the subsystems (here, both subsystems 
are assumed to have equal dimensions). Hence, when 
there is notable amplification, then the number of Fock states 
to be taken into account for sufficient numerical accuracy 
drastically increases. In contrast to absorbing media, where
the dimension of the Hilbert space of the relevant modes  
is bounded by the total number 
of input photons, such a bound does not exist for amplifying media. 

Nevertheless, for entangled Gaussian states an upper bound of the
gain can be determined such that the amplified system is still 
not separable. Let us consider, e.g., the two-mode squeezed vacuum 
(\ref{sqvac2}) and assume that the two modes travel through amplifying
devices at zero temperature. The Wigner function of the
two-mode squeezed vacuum is a Gaussian
\begin{equation}
\label{amp2}
W(\mbb{\xi}) = \left(4\pi^2 \sqrt{\det \mbb{V}}\right)^{-1}
\exp\!\left\{ -{\textstyle\frac{1}{2}} 
\mbb{\xi}^T \mbb{V}^{-1} \mbb{\xi} \right\}.
\end{equation}
Here, $\mbb{\xi}$ is a four-vector whose elements are
$q_1,p_1,q_2,p_2$, and $\mbb{V}$ is the $4\times 4$ variance matrix
\begin{equation}
\label{amp3}
\mbb{V} = \left( \begin{array}{cc}
{\bf X} & {\bf Z} \\ {\bf Z}^T & {\bf Y} 
\end{array} \right) \,.
\end{equation}
The variance matrix (\ref{amp3}) can be written in the form
\begin{equation}
\label{amp7}
{\bf V} = \left( \begin{array}{cccc}
c/2&0&-s/2&0\\0&c/2&0&s/2\\-s/2&0&c/2&0\\0&s/2&0&c/2
\end{array} \right)
\end{equation}
[$c$ $\!=$ $\!\cosh 2\zeta$, $s$ $\!=$ $\!\sinh 2\zeta$].
Using the input-output relations (\ref{2.4}) for amplifying devices, 
we can easily transform the input-state
variance matrix (\ref{amp7}) to obtain the 
output-state variance matrix
\begin{equation}
\label{amp7a}
{\bf V} = \left( \begin{array}{cccc}
x&0&Z_{11}&Z_{12}\\
0&x&Z_{21}&Z_{22}\\
Z_{11}&Z_{21}&y&0\\
Z_{12}&Z_{22}&0&y
\end{array} \right),
\end{equation}
where
\begin{equation}
x = \textstyle\frac{1}{2}c |T_1|^2 +\textstyle\frac{1}{2} |R_1|^2
+\textstyle\frac{1}{2} \left( |T_1|^2 + |R_1|^2 -1 \right) ,
\end{equation}
\begin{equation}
y = \textstyle\frac{1}{2}c |T_2|^2 +\textstyle\frac{1}{2} |R_2|^2
+\textstyle\frac{1}{2} \left( |T_2|^2 + |R_2|^2 -1 \right) ,
\end{equation}
\begin{eqnarray}
Z_{11} = -Z_{22} &=& -\textstyle\frac{1}{2}s
\,{\rm Re}\left( T_1 T_2 \right) , \\ 
Z_{12} = Z_{21} &=& -\textstyle\frac{1}{2}s
\,{\rm Im}\left( T_1 T_2 \right) .
\end{eqnarray}
Let us consider equal devices, so that $T_1$ $\!=$ $\!T_2$ $\!=T$
and $R_1$ $\!=$ $\!R_2$ $\!=R$. The Peres--Horodecki criterion
\cite{Simon00} 
\begin{eqnarray}
\label{amp4}
\lefteqn{
\det{\bf X} \det{\bf Y} 
+ \left( {\textstyle\frac{1}{4}}-|\det{\bf Z}| \right)^2
- {\rm Tr}\,\big( {\bf XJZJYJZ}^T{\bf J}\big)
} 
\nonumber \\ &&\hspace{6ex} 
\ge {\textstyle\frac{1}{4}} 
(\det{\bf X} +\det{\bf Y}) \,, \qquad
{\bf J} = \left( 
\begin{array}{cc} 0 & 1 \\ -1 & 0 
\end{array} \right)
\end{eqnarray}
then tells us that for 
\begin{equation}
\label{amp8}
|T|^2 = \frac{2\left(1-|R|^2\right)}{1+{\rm e}^{-2|\zeta|}}
\end{equation}
the boundary between separability and nonseparability is reached.
In particular for zero reflection ($R$ $\!=$ $\!0$), Eq.~(\ref{amp8})
reveals that the upper limit of the gain 
\mbox{$g$ $\!=$ $\!|T|^2$ $\!-$ $\!1$ $\!\ge$ $\!0$}
for which nonseparability changes to separability is simply given
by the squeezing parameter $|q|$,
\begin{equation}
\label{amp9}
g = |q| = \tanh |\zeta| \,.
\end{equation}
An obvious consequence of Eq.~(\ref{amp9}) is that entanglement cannot 
be produced from the vacuum by amplification. Since for the vacuum 
the squeezing parameter has to be set equal to zero, $q$ $\!=$ $\!0$,
from Eq.~(\ref{amp9}) it follows that any nonvanishing gain $g$ must 
necessarily lead to a separable state.


\section{Summary and conclusions}
\label{Conclusion}

We have studied the problem of quantum-state transformation at
absorbing and amplifying dielectric four-port devices, without
making use of any replacement schemes. We instead express the
input--output relations in terms of the actually observed
quantities as obtained from the quantized Maxwell
field in the presence of arbitrary causal (linear) media.  
After extending the basic formulas recently developed
for absorbing media to amplifying media, we have applied the
theory to some problems typically considered in quantum 
information processing. 

In particular, we have considered both the amount 
of entanglement that is realized when nonclassical 
light is combined through a lossy beam splitter and the
entanglement degradation when entangled light propagates
through lossy media. We have based our analysis on the
quantum relative entropy as a measure of entanglement. 
The calculation of the entanglement of a mixed quantum state  
typically observed for absorbing media needs comparing the state 
with all separable states in order to find that separable state 
which is closest to the state under consideration. Since
the effort drastically increases with the dimension of the
Hilbert space, we have restricted our attention to low-dimensional 
quantum states in the numerical calculation. 

The numerical results show
that the Bell-type states $|\Psi_n^\pm\rangle$, Eq.~(\ref{trans0}),
are more robust against decoherence than the states 
$|\Phi_n^\pm\rangle$, Eq.~(\ref{trans0a}) ($n$ $\!=$ $\!1,2$).
The estimation of an upper bound of entanglement for 
arbitrary number $n$ of photons in each of the two entangled
modes shows that with increasing $n$ the 
characteristic length of entanglement degradation decreases as 
$L/n$ at least, where $L$ is the absorption length
according to the Lambert--Beer law.

So far we have considered either purely absorbing or purely
amplifying media. In practice, the two effects can occur
simultaneously. Essentially, there are two ways to deal with  
this problem. One way is to treat amplifiers with absorption
as cascading amplifying and absorbing devices. Another way
is to go back to the underlying quantized Maxwell equations
with the aim to develop a more specific approach to the problem.


\acknowledgements

S.S. gratefully acknowledges support by the Adam Haker
Fonds. S.S. also likes to thank M.~Fleischhauer for helpful
discussions concerning electromagnetically induced transparent media.
This work was supported by the Deutsche Forschungsgemeinschaft.


\appendix

\section{Density matrix for Gaussian Wigner function}

As mentioned in Sec.~\ref{uniop} in the case of amplifying media only
symmetric operator ordering is preserved, and 
hence the Wigner function (\ref{3.10}) is suited to the 
state description. For the sake of transparency we will
restrict our attention to a single-frequency component
[i.e., a (qua\-si-)mo\-nochromatic field in a sufficiently
small frequency interval $\Delta \omega$ \cite{{Knoll99}}).
The extension to a multifrequency field is straightforward. 
When the input field is prepared in a Fock state $|p,q\rangle$
and the device in the ground state $|0,0\rangle$, so that
the overall input state is $|p,q,0,0\rangle$, then the input
Wigner function reads as
\begin{eqnarray}
\label{A1}
\lefteqn{
W_{\rm in}
(\mbb{\alpha}, \mbb{\alpha}^\ast )
= \left( \frac{2}{\pi} \right)^4 (-1)^{p+q} \,
{\rm e}^{-2(|g_1|^2+|g_2|^2)} 
}
\nonumber \\ && \hspace{6ex} \times\,
L_p(4|a_1|^2) \,L_q(4|a_2|^2) \,{\rm e}^{-2(|a_1|^2+|a_2|^2)} 
\end{eqnarray}
with $L_n(x)$ being the Laguerre polynomial
\begin{equation}
\label{A2}
L_n(x) = \sum\limits_{m=0}^n (-1)^m {n \choose n-m} \frac{x^m}{m!} \,.
\end{equation}
We now apply Eq.~(\ref{3.10}), making the substitutions 
according
\begin{eqnarray}
\label{A4}
{\bf a} &\to& {\bf T}^+ {\bf a} - {\bf T}^+ {\bf C}^{-1} {\bf S} \,
{\bf g}^\ast ,
\\
{\bf a}^\ast &\to& {\bf T}^T {\bf a}^\ast - {\bf T}^T
\left[{\bf C}^T\right]^{-1} {\bf S}^T {\bf g} ,
\\
{\bf g} &\to& -{\bf A}^T {\bf a}^\ast + {\bf A}^T
\left[{\bf S}^T\right]^{-1} {\bf C}^T {\bf g} ,
\\
{\bf g}^\ast &\to& -{\bf A}^+ {\bf a} + {\bf A}^+ {\bf S}^{-1} {\bf C} 
\, {\bf g}^\ast .
\end{eqnarray}
Finally, we integrate over the device variables $g_i$ to obtain
the Wigner function of the outgoing field.
Introducing the matrix $K_{ii'}$ $\!=$ $\!\delta_{ii'}k_i$
and employing the formula
\begin{equation}
\label{A3}
4|a|^2 {\rm e}^{-2|a|^2} = \frac{\partial}{\partial k}
\left. {\rm e}^{-2|a|^2+4k|a|^2} \right|_{k=0} ,
\end{equation}
we derive
\begin{eqnarray}
\label{A10}
\lefteqn{
W_{\rm out}^{(F)}
({\bf a},{\bf a}^\ast) 
}
\nonumber \\ && \hspace{1ex}
= \left( \frac{2}{\pi} \right)^2
\sum\limits_{h=0}^p\sum\limits_{l=0}^q
\frac{(-1)^{h+p}}{h!} {p \choose h}
\frac{(-1)^{l+q}}{l!} {q \choose l}
\frac{\partial^h}{\partial k_1^h}
\frac{\partial^l}{\partial k_2^l}
\nonumber \\ && \hspace{1ex}\times\,
\left. \frac{\exp\!\left\{-2({\bf a}^\ast)^T
[{\bf N}-{\bf B}^T ({\bf D}^T)^{-1} {\bf B}^\ast] {\bf a}\right\}}
{\det {\bf D}} 
\right|_{k_1=k_2=0}\!, 
\end{eqnarray}
where the abbreviations
\begin{eqnarray}
\label{A11}
{\bf N} &=& 2{\bf T}{\bf T}^+ -{\bf I}-2{\bf T}{\bf K}{\bf T}^+ , 
\\
{\bf B} &=& 2{\bf S}^T{\bf C}^T-2{\bf S}^\ast {\bf C}^{\ast\,-1}{\bf T}
{\bf K} {\bf T}^T , 
\\
{\bf D} &=& 2 {\bf T}^\ast {\bf T}^T -{\bf I} -2 {\bf S}^\ast {\bf
C}^{\ast\,-1} {\bf T}^\ast {\bf K} {\bf T}^T {\bf C}^{T\,-1} {\bf S}^T.
\end{eqnarray}
have been used.

In order to calculate from the Wigner function the density 
operator, we make use of the relation \cite{Cahill69}
\begin{equation}
\label{A12}
\hat{\varrho}^{(F)}_{\rm out} = \pi^2 \int {\rm d}^2{\bf a} \,
W^{(F)}_{\rm out}({\bf a},{\bf a}^\ast)
\,\hat{\delta}({\bf a}-\hat{\bf a}) ,
\end{equation}
where
\begin{equation}
\label{A13}
\hat{\delta}({\bf a}-\hat{\bf a}) 
= \frac{1}{\pi^4} \int {\rm d}^2{\bf b} 
\, \hat{D}({\bf b}) \,{\rm e}^{{\bf a}^T{\bf b}^\ast
-{\bf b}^T{\bf a}^\ast} ,
\end{equation}
with $\hat{D}({\bf b})$ being the two-mode coherent displacement
operator. For notational convenience we introduce the
abbreviation notation
\begin{eqnarray}
\label{A13a}
\lefteqn{
{\cal D} \{\ldots\}
= \sum\limits_{h=0}^p\sum\limits_{l=0}^q
\Bigg[
\frac{(-1)^{h+p}}{h!} {p \choose h}
\frac{(-1)^{l+q}}{l!} {q \choose l}
}
\nonumber\\&&\hspace{15ex}\times\,
\left.\frac{\partial^h}{\partial k_1^h}
\frac{\partial^l}{\partial k_2^l}
\Big\{\ldots\Big\} \Bigg] \right|_{k_1=k_2=0}.
\end{eqnarray}
Substitution of Eq.~(\ref{A10}) into Eq.~(\ref{A12}) yields
\begin{eqnarray}
\label{A14}
\lefteqn{
\hat{\varrho}^{(F)}_{\rm out} = 
{\cal D}\bigg\{
\frac{4}{\pi^4 \det {\bf D}}
\int \big[ {\rm d}^2{\bf a} \,{\rm d}^2{\bf b} \,\hat{D}({\bf b})
}
\nonumber \\ &&\hspace{8ex}\times\,
{\rm exp}\!\left(-2 {\bf a}^+ {\bf M} {\bf a} + {\bf a}^T
{\bf b}^\ast -{\bf b}^T {\bf a}^\ast \right) \big]
\bigg\},
\end{eqnarray}
where ${\bf M}$ $\!\equiv$ $\!{\bf N}$ $\!-$ $\!{\bf B}^T 
({\bf D}^T)^{-1} {\bf B}^+$.

Using the Fock-state representation of the (single-mode) coherent
displacement operator \cite{Cahill69},
\begin{equation}
\label{A15}
\langle m | \hat{D}(b) | n \rangle =
\sqrt{\frac{n!}{m!}} \,b^{m-n} {\rm e}^{-|b|^2/2}
L_{n}^{(m-n)}(|b|^2) 
\end{equation}
[$L_{n}^{m}(x)$, associated Laguerre polynomial],
we can calculate the density matrix in the Fock basis.
Performing the ${\bf a}$ integrals in Eq.~(\ref{A14}), 
we derive
\begin{eqnarray}
\label{A17}
\lefteqn{
\langle m_1,m_2| \hat{\varrho}^{(F)}_{\rm out} |n_1,n_2\rangle = 
{\cal D}\bigg\{
\frac{1}{\pi^2 \det {\bf DM}}
}
\nonumber\\&&\hspace{2ex}\times
\sum\limits_{n_1,n_2,m_1,m_2} 
\sqrt{\frac{n_1!n_2!}{m_1!m_2!}}
\int 
r_1{\rm d}r_1\,r_2{\rm d}r_2\,{\rm d}\varphi_1\,
{\rm d}\varphi_2 \, 
\nonumber\\&&\hspace{2ex}\times\,
r_1^{m_1-n_1} r_2^{m_2-n_2}\, 
{\rm exp}\!\left[ -\frac{1}{2} r_1^2 \left( 1+\frac{M_{22}}{\det{\bf M}} 
\right)\right.   
\nonumber \\ &&\hspace{2ex} 
\left.-\frac{1}{2} r_2^2 \left( 1+\frac{M_{11}}{\det{\bf M}} \right)
+\frac{|M_{12}|}{\det{\bf M}} r_1r_2 \cos(\Theta+\varphi_2-\varphi_1)
\right]   
\nonumber \\ &&\hspace{2ex}\times\,
{\rm e}^{i\varphi_1(m_1-n_1)+i\varphi_2(m_2-n_2)}\,
L_{n_1}^{(m_1-n_1)}(r_1^2) 
\nonumber \\ &&\hspace{2ex}\times\,
L_{n_2}^{(m_2-n_2)}(r_2^2) 
\bigg\} ,
\end{eqnarray}
where we have used the notation $b_i$ $\!=$ $\!r_ie^{i\varphi_i}$,
and $M_{12}$ $\!=$ $\!|M_{12}| {\rm e}^{i\Theta}$.
Recalling the definition of the modified Bessel functions,
we perform the angular integrals to obtain
\begin{eqnarray}
\label{A18}
\lefteqn{
\langle m_1,m_2| \hat{\varrho}^{(F)}_{\rm out} |n_1,n_2\rangle = 
}
\nonumber \\&&\hspace{2ex}
{\cal D}\bigg\{
\frac{1}{\det {\bf DM}}
\sum\limits_{n_1,n_2,m_1,m_2} \sqrt{\frac{n_1!n_2!}{m_1!m_2!}}
\nonumber\\&&\hspace{2ex}\times\,
{\rm e}^{-i\Theta(m_2-n_2)}
\,\delta_{m_1-n_1+m_2-n_2,0} 
\nonumber \\ &&\hspace{2ex} \times
\int\limits_0^\infty 
{\rm d}x_1\,{\rm d}x_2 \,
{\rm exp}\!\left[
-{\textstyle\frac{1}{2}} x_1 \left( 1+\frac{M_{22}}{\det{\bf M}} \right)
\right.
\nonumber\\&&\hspace{2ex} 
\left.-{\textstyle\frac{1}{2}} x_2 \left( 1+\frac{M_{11}}{\det{\bf M}} \right)
\right] 
I_{m_2-n_2}\!\left( \frac{|M_{12}|}{\det{\bf M}} \sqrt{x_1x_2} \right)
\nonumber \\ && \hspace{2ex}\times\,
x_1^{(m_1-n_1)/2} x_2^{(m_2-n_2)/2} L_{n_1}^{(m_1-n_1)}(x_1) \,
\nonumber \\ && \hspace{2ex}\times\,
L_{n_2}^{(m_2-n_2)}(x_2) 
\bigg\}
\end{eqnarray}
($x_i$ $\!=$ $\!r_i^2$).
The $x_2$ integral is performed by means of the formula (2.19.12.6) in
\cite{Prudnikov}, which gives (for $m_2$ $\!\ge$ $\!n_2$)
\begin{eqnarray}
\label{A19}
\lefteqn{
\langle m_1,m_2| \hat{\varrho}^{(F)}_{\rm out} |n_1,n_2\rangle =
}
\nonumber\\&&\hspace{2ex}
{\cal D}\bigg\{
\frac{2}{\det {\bf D}}
\!\sum\limits_{n_1,n_2,m_1,m_2}\! \sqrt{\frac{n_1!n_2!}{m_1!m_2!}}
\left( M_{12}^\ast \right)^{m_2-n_2} 
\nonumber \\ &&\hspace{2ex} \times\,
\delta_{m_1-n_1+m_2-n_2,0}\,
\frac{(M_{11}-\det{\bf M})^{n_2}}{(M_{11}+\det{\bf M})^{m_2+1}}
\nonumber \\ &&\hspace{2ex} \times
\int\limits_0^\infty {\rm d}x_1 \, {\rm exp}\!\left[
-\frac{1}{2} x_1 \left( 1+\frac{1+M_{22}}{M_{11}+\det{\bf M}} \right)
\right]
\nonumber \\ &&\hspace{2ex} \times\,
L_{n_1}^{(m_1-n_1)}(x_1) \, L_{n_2}^{(m_2-n_2)}\!\left(
\frac{|M_{12}|^2}{M_{11}^2-(\det{\bf M})^2} x_1 \right)
\bigg\} .
\nonumber \\
\end{eqnarray}
Finally, the $x_1$ integral is performed by expanding the associated
Laguerre polynomials into power series \cite{Gradstein}. The result is
\begin{eqnarray}
\label{A20}
\lefteqn{
\langle m_1,m_2| \hat{\varrho}^{(F)}_{\rm out} |n_1,n_2\rangle = 
}
\nonumber\\&&\hspace{2ex}
{\cal D}\bigg\{
\frac{2}{\det {\bf D}}
\!\sum\limits_{n_1,n_2,m_1,m_2}\! \sqrt{\frac{n_1!n_2!}{m_1!m_2!}} 
\nonumber \\ &&\hspace{2ex} \times\,
\delta_{m_1-n_1+m_2-n_2,0}\,
\frac{(M_{11}-\det{\bf M})^{n_2}}{(M_{11}+\det{\bf M})^{m_2+1}}
\nonumber \\ &&\hspace{2ex} \times \,
\left(M_{12}^\ast \right)^{m_2-n_2}
{m_1 \choose n_1}
\sum\limits_{k=0}^{n_2}  \frac{c^k}{a^{k+1}} {m_2 \choose
n_2-k}
\nonumber \\ &&\hspace{2ex} \times\,
{}_2F_1\!\left( k+1,-n_1;m_1-n_1+1,\frac{1}{a}\right) \bigg\} \,, 
\end{eqnarray}
where
\begin{eqnarray}
\label{A21}
a &=& \frac{1+M_{11}+M_{22}+\det{\bf M}}{2(M_{11}+\det{\bf M})} \,,
\\
c &=& \frac{|M_{12}|^2}{(\det{\bf M})^2-M_{11}^2} \,.
\end{eqnarray}

Integrating Eq.~(\ref{A10}) over the phase space of one mode
of the outgoing field yields the Wigner function of the quantum 
state of the other mode
\begin{eqnarray}
\label{A22}
\lefteqn{
W_{\rm out}^{(F)}
(a_i,a_i^\ast)
}  
\nonumber \\ && \hspace{2ex} 
= \,\frac{2}{\pi} 
\sum\limits_{h=0}^p \sum\limits_{l=0}^q
\frac{(-1)^{h+p}}{h!} {p \choose h}
\frac{(-1)^{l+q}}{l!} {q \choose l}
\nonumber \\ && \hspace{2ex} \times \left.
\frac{\partial^h}{\partial k_1^h}\frac{\partial^l}{\partial k_2^l}
\frac{\det {\bf E}}{E_{ii}
\det {\bf D}} \,{\rm e}^{ -2|a_i|^2/E_{ii}}
\right|_{k_1=k_2=0} .
\end{eqnarray}
(${\bf E}$ $\!=$ $\!{\bf M}^{-1}$).
This Wigner function is equivalent to the density matrix
in the Fock basis
\begin{eqnarray}
\label{A23}
\hat{\varrho}_{{\rm out}\,i}^{(F)} &=& \sum\limits_{n=0}^\infty 
\Bigg[ \sum\limits_{h=0}^p \sum\limits_{l=0}^q 
\frac{(-1)^{h+l+p+q}}{h!l!} {p \choose h}{q \choose l} \nonumber
\\ && \hspace{-8ex} \times 
\frac{\partial^{h+l}}{\partial k_1^h \partial k_2^l} 
\frac{\det {\bf E}}{\det {\bf D}} \frac{2}{E_{ii}+1} \left(
\frac{E_{ii}-1}{E_{ii}+1} \right)^n \Bigg]_{k_1=k_2=0} \hspace{-3ex}
|n\rangle\langle n| .
\end{eqnarray}


\end{document}